\documentstyle[aps,prl,epsfig,multicol]{revtex}
 
\begin{document}
\draft
\title{
Transition from molecular complex to quantum solvation in
$^4$He$_N$OCS}
\author{F. Paesani$^{1,2}$, A. Viel$^{1}$\footnote{
current address: Lehrstuhl f\"ur Theoretische Chemie, Technische Universit\"at 
M\"unchen, 
85747 Garching, Germany},
F. A. Gianturco$^{2}$, and 
K. B. Whaley$^1$}
\address{$^1$Department of Chemistry and Kenneth S. Pitzer Center for
Theoretical Chemistry, University of
California, Berkeley, CA 94704 \\
$^2$ Department of Chemistry and INFM, University of Rome "La Sapienza", 
Citt\`a Universitaria, 00185 Rome, Italy 
}
\date{July 18, 2002}
\maketitle

\begin{abstract}
We present quantum calculations of the rotational energy levels and
spectroscopic rotational constants of the
linear OCS molecule in variable size clusters of $^4$He,
using spectral evolution quantum Monte Carlo methods that allow excited states
to be accessed without nodal constraints.
The rotational constants of
OCS are found to decrease monotonically from the gas phase value as the number 
of helium
atoms increases to $N=6$, after which the average constant
increases to saturation at a value
in excellent agreement with experimental measurements
made on significantly larger clusters ($N \geq 1000$).  
The minimum is shown to indicate a transition from a molecular
complex to a quantum solvated molecule, with the former characterized by floppy 
but near
rigid behavior, while the latter is characterized by non-zero permutation 
exchanges 
and a smaller extent of rigid coupling. 
\end{abstract}


\begin{multicols}{2}

\narrowtext
Helium droplets offer a unique opportunity to immerse molecules into
a superfluid.
Spectroscopic experiments on molecules embedded in these droplets have yielded 
an array
of measurements that provide on the one hand access to elementary 
excitations of the finite quantum liquid, and on the other
hand information about the novel solvation dynamics encountered in a 
superfluid \cite{seetoennies01}.
The rotational dynamics of the linear OCS molecule has played a key role
in these spectroscopic studies.
When solvated by the bosonic $^4$He isotope at temperatures
$T \leq 0.5$ K, infra-red
spectra of OCS show rotational fine
structure that is consistent with a free rotation accompanied by an 
increased molecular moment of inertia, while
no fine structure is observed when OCS is solvated by fermionic $^3$He 
clusters~\cite{grebenev98}. 
Quantum Monte Carlo (QMC) calculations have confirmed 
that a spectrum of free rotational states with modified moment of inertia
will result from bosonic solvation \cite{seekwon00}. Such free 
rotation is not just the result of solvation by a fluid interacting with 
weak van der Waals forces.  It
has been shown to be consistent with a negligible transfer of angular momentum 
from the molecule to relative motion of the solvating helium when the latter 
possesses 
Bose permutation symmetry \cite{babichenko99,seekwon00}. 
In contrast, 
efficient coupling to single-particle excitations in 
the fermionic $^3$He droplets 
has been shown to considerably reduce the lifetime of the rotational 
excitations~\cite{babichenko99}, accounting for the absence of rotational 
spectral 
transitions.

Calculations of excited states afforded by QMC yield not only a direct route
to calculation of spectroscopic constants for molecules solvated in helium, but 
they also
provide critical insight
into the nature of these rotational excitations. They thereby allow 
detailed
microscopic analysis of the quantum coupling between a molecule and its helium
environment. 
Direct calculations of rotational energy levels have been made within dynamical 
approximations such as
the quasi-adiabatic rigid coupling approximation~\cite{quack91,seekwon00} and
fixed node approximations~\cite{lee99,viel01}. However, 
in order to provide a complete understanding of the underlying quantum dynamics,
it is necessary to use a 
theoretical approach that allows {\em exact} calculation of excited states, 
devoid of any 
dynamical or nodal approximations. 
This is possible with
the projection operator imaginary time spectral evolution
(POITSE) approach, a diffusion-Monte-Carlo-based methodology that 
has recently been
applied to calculation of rotational excitations of the linear HCN molecule in 
helium clusters~\cite{viel01}. 
For HCN, a considerably lighter
molecule than OCS, these results 
showed a complex coupling between molecular and helium degrees of freedom 
that was attributed to 
the high zero point energy of the He-HCN system.
In contrast, for the heavier octahedral molecule SF$_6$, rotational
constants derived from fixed node
calculations employing the free molecule nodal surfaces showed
excellent agreement with experimental measurements~\cite{lee99}. 

In this letter we investigate the rotational motion of the OCS molecule inside 
$^4$He clusters,
with direct calculations of rotational energy levels by POITSE. 
Analysis of the results in terms of rigid coupling approximations and of 
permutation exchanges visible in related finite temperature calculations allow 
for the 
first time a clear transition
from a {\em molecular complex} to a molecule {\em solvated by a quantum liquid} 
to be identified as a
function of cluster size $N$. 
OCS is intermediate in mass between HCN and SF$_6$. 
Comparison of the rotational behavior of OCS with that of HCN and of SF$_6$ 
provides
key insights into the quantum coupling of the molecular rotational motion with 
the 
solvating helium degrees of freedom, as a function of both molecular mass and 
symmetry. 

Our calculations employ the Projector Operator Imaginary Time Spectral Evolution 
(POITSE) method for calculation of excited states\cite{blume97}.
In this scheme, excited state energies are extracted
from the two-sided inverse Laplace transform of an imaginary time
correlation function $\tilde{\kappa}(\tau)$ that is  
computed by combining a multi-dimensional Monte Carlo
integration with Diffusion Monte Carlo sidewalks.
The decay of the correlation
$\tilde{\kappa}(\tau)$ contains information about energy differences $E_f-E_0$,
where $E_0$ is the ground state energy and $E_f$ an excited state
energy level. 
The POITSE approach is ideal for calculation of excitations in many-particle 
systems when
estimates of excitation functions that take the ground state to the desired 
excited
states are available. A suitable correlation function is then
\begin{equation}
\tilde{\kappa}(\tau)=\frac
    {\langle \Psi_T \mid \hat{A} \exp[-(\hat{H}-E_0)\tau]\hat{A}^{\dag} \mid 
\Psi_T \rangle}
    {\langle \Psi_T \mid  \exp[-(\hat{H}-E_0)\tau] \mid \Psi_T \rangle},
\label{eq:time_correl_func}
\end{equation}
where $\hat{A}$ is a local operator that projects from a trial
function $\mid\Psi_T\rangle$ approximating the ground state $\mid\Psi_0\rangle$. 
The resulting initial 
state $\hat{A}^{\dag} \mid\Psi_T\rangle$
is usually not an eigenfunction of ${\hat H}$, but 
time evolution under the action of the full Hamiltonian will ensure that all 
eigen components $\mid\Psi_f\rangle$ that overlap with this state
will contribute to the correlation function. 
Inverse Laplace transform of $\tilde{\kappa}(\tau)$, performed by the maximum 
entropy 
method as described in Ref.~\onlinecite{blume97}, results in the 
spectral function 
\begin{equation}
\kappa(\omega) = \sum_f \mid \langle \Psi_T \mid \hat{A} \mid \Psi_f \rangle 
\mid^2
                 \delta(E_0-E_f+\omega),
\label{eq:spect_func}
\end{equation}
from which all excitation energies accessed by $\hat{A}$ may be extracted as the 
peak locations. When $\mid\Psi_T\rangle \equiv \mid\Psi_0\rangle$, exact 
energies are 
obtained, while otherwise
some trial function bias may exist \cite{blume97}.

Our approach to extract the rotational excitations of interest
is to take a free molecular projector acting on the
molecule-helium cluster ground state. Since the molecule-helium coupling is 
weak, 
this provides a good starting point for calculation of the
imaginary time correlation function and subsequent spectral transformation.
For the trial function $\Psi_T$ we employ 
variationally optimized ground state functions of the usual generalized product 
form, namely
containing pairwise correlations between all components of
the cluster, {\em i.e.}, isotropic He-He and anisotropic He-OCS terms.  
The He-He correlations are those used in 
Ref.~\onlinecite{paesani01c}, while the He-OCS correlations take the form
$\chi(R,\theta;\{p_i\})= \exp\{p_0R^{p_1}+p_2[1+p_3\cos(\theta-p_4)\ln R] 
  +p_5R^2\cos^2(\theta-p_4)e^{(p_6-p_7R)}\}$
where $R, \theta$ are the Jacobi coordinates describing the location of the He 
atom
relative to the linear OCS molecule~\cite{paesani01c}. 
We consider here projectors $\hat{A}$ proportional to the
molecular Wigner functions $D_{mk}^{j}$ in a space-fixed frame.
Zeros of $D_{mk}^{j}$ constitute free molecule nodal surfaces for a symmetric 
top~\cite{zare}.
We focus here in particular 
on $j=1$, $m=k=0$, a linear rotor eigenfunction for which 
$\hat{A}$=$\cos\beta$ 
is a function only of the second Euler
angle of the molecule 
specifying the orientation of the 
molecular frame in the (arbitrary) space-fixed frame. 
$\hat{A}$ accesses states in which 
the total angular momentum $J$
is carried primarily by the OCS, {\it i.e.}, 
the molecular angular momentum $j$ is a quasi-good quantum number
and there is negligible angular momentum $l$ in the relative
helium motion (${\bf J} = {\bf j} + {\bf l}$).
Our choice of $D_{00}^{1}$ is motivated
by analysis of the calculated rotational spectrum for the He-OCS complex ($N=1$) 
\cite{gianturco00b} which shows that this is the primary molecular function
contributing to the $J_{K_pK_o}$=1$_{01}$ rotational state of the complex.  (We 
employ
here the notation $J, K_p, K_o$ for rotational states of an
asymmetric rotor~\cite{zare}). 

We emphasize that the projector ${\hat A}$
provides only an initial guess for the nodal structure of the
excitations, and that these automatically adjust to their true values during the 
calculation. That these are indeed subtly different from the free nodal surfaces 
for OCS is evident from the fact that for $N=1$, a 5-dimensional system where 
exact calculations are possible using {\it e.g.}, the BOUND 
program~\cite{comp_bound}, fixed node calculations carried out with nodal 
constraints imposed by the Wigner function do not show good agreement with the 
exact calculations, overestimating the exact rotational excitation energy 
for OCS-He (0.307 cm$^{-1}$) by 0.07$\pm$0.01 cm$^{-1}$, {\it i.e.}, 22(3)\%.  In contrast, POITSE gives the 
exact energy to within a statistical error of 0.7\%, {\it i.e.}, 0.309$\pm$0.002 cm$^{-1}$. 
This behavior contrasts with the high accuracy of the free molecule fixed node approximation found 
for the heavier 
SF$_6$-He complex.~\cite{lee99} 

The total molecule-cluster interaction potential
is the sum of all pairwise contributions (He-He and OCS-He).
We have employed two recent He-OCS potentials~\cite{gianturco00b,higgins99} 
and find equivalent results with these.  Unless otherwise stated, all 
calculations reported here
are made with the {\em ab initio} HHDSD He-OCS potential energy surface 
\cite{gianturco00b}
and the HFD-B He-He potential of Ref.~\cite{aziz87}. 
The imaginary time evolution is performed with the rotational importance-sampled
rigid body Diffusion Monte Carlo algorithm of Ref.~\cite{viel02}. 
Initial ensembles of 1000 walkers distributed according to $|\Psi_T|^2$ were
propagated for 30,000 
steps of $\Delta\tau$=50~a.u. using
the mixed branching/weight algorithm of Ref.~\cite{huang02_short}. Typically 
1000-2000 independent
decays were required to produce a converged spectrum $\kappa(\omega)$, with the 
largest
size ($N = 20$) requiring the most decays. Statistical error bars in the 
excitation 
energies were
estimated using a Gaussian approximation, according to the procedure described 
in 
Ref.~\onlinecite{sivia}. 

POITSE calculations with the Wigner projector ${\hat A}$ consistently produced 
only one 
peak in the 
$\kappa(\omega)$ spectrum, for all sizes $N \leq 20$. This clean single-valued 
behavior 
of the free molecule
projector for OCS contrasts with the behavior seen earlier for the lighter HCN 
molecule,
where the same projector leads to multiple excited states~\cite{viel01}.  
Single-valued free molecule projectors suggest that ${\hat A}|\Psi_T\rangle$ is 
very
similar to the true eigenstate.
However, free or quasi-free rotations do 
not prohibit effective rigid coupling of some solvating helium density to the 
molecular
rotation, leading to a renormalization of the molecular moment of inertia  
similar to that found in the microscopic two-fluid theory~\cite{seekwon00} and 
in perturbative analysis~\cite{babichenko99}.

\begin{figure}[h]
\epsfig{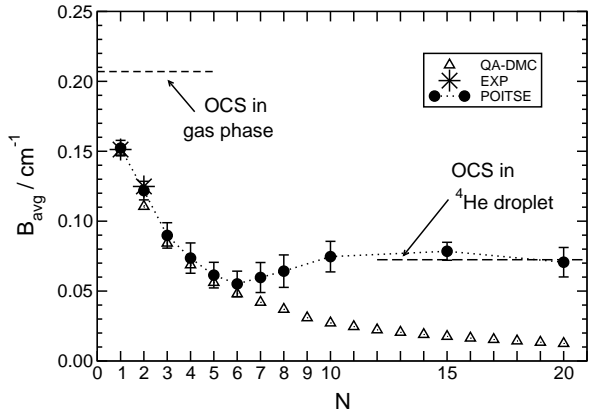}
\vspace{0.5cm}
\caption[]{Effective rotational constant $B_{avg}$ for OCS in $^4$He$_N$,
as a function of cluster size
$N$.  Circles show POITSE results obtained from 
cluster rotational state energies generated by projectors 
$\hat{A}=\cos(\beta)$ (see text).
Triangles show quasi-adiabatic rigid coupling 
estimates~\onlinecite{seepaesani01b}, asterisks correspond to the experimental 
values of $B_{avg}$
for $N$=1,2~\cite{higgins99,xu01}.  The latter are seen to be coincident 
with the POITSE results (circles).
Gas phase and experimental values in large $^4$He droplets are taken from 
Ref.\cite{grebenev00b}.}
\label{fig1}
\end{figure}

When interpreted with an asymmetric top Hamiltonian~\cite{zare}, the cluster
rotational excited state energy derived from ${\hat A}$, 
$E(J_{K_p,K_o} =1_{01})$, yields an effective rotational 
constant $B_{avg}$. When close to the limit of a symmetric top, $B_{avg}$ is
equal to either $(B+C)/2$ (prolate top) or $(A+B)/2$ (oblate top).  
Fig.~\ref{fig1} shows the main result of this paper, namely the behavior of 
$B_{avg}$ 
obtained from the POITSE calculations, as a function of 
cluster size $N$.   
Corresponding experimental values of $B$ for the free (gas 
phase) OCS molecule and for OCS in large helium droplets 
($N \geq 1000$) where the spectrum
was fit to a linear rotor ($B_{avg} \equiv B$) are shown from 
Ref.~\cite{grebenev00b}. 
For the smallest clusters 
$N=1$ \cite{higgins99} and $N=2$ \cite{xu01} we show the experimentally measured 
values of $B_{avg}$.

The POITSE results show a monotonic decrease of the excited state energy up to 
$N=6$, 
followed by a small continuous increase to saturation at a value 
$B_{avg}=0.07(2)$ cm$^{-1}$ that is in excellent agreement with the
experimentally measured value $B=0.0732(3)$ cm$^{-1}$ in large 
droplets\cite{grebenev00b}.
Excellent agreement of $B_{avg}$ with the corresponding 
experimental values $B_{avg}$ for small clusters is also achieved for the two 
sizes 
$N$=1, 2 for which this experimental data is available \cite{higgins99,xu01}.
Thus the POITSE calculations are able to bridge the gap between small van
der Waals clusters
and large droplets, achieving high accuracy in both regimes. 
The overall behavior of $B_{avg}$ for OCS shows some similarities with SF$_6$ 
and with 
HCN, but also exhibits critical differences from these molecules.  For the more
symmetrical and heavier SF$_6$ molecule
the saturation value is also reached before
the first solvation shell is completed ($N\simeq 8$) but negligible overshoot
is seen~\cite{lee99}.  
For the lighter HCN, also linear, a large overshoot is seen but the
rotational constant does not reach saturation before completion of the first 
solvation 
shell~\cite{viel01}.

A deeper understanding of the size dependent quantum coupling within these 
rotational 
excitations in the bosonic $^4$He cluster is provided by analysis in terms of 
both rigid 
coupling models and permutation exchange distributions.
The triangle symbols in Fig.~\ref{fig1} show the values of $B_{avg}$ 
obtained using the quasi-adiabatic rigid coupling 
approach \cite{quack91,seekwon00} implemented with correlated 
sampling~\cite{seepaesani01b}. 
The POITSE results are very close to the rigid coupling 
estimates for $N\leq6$, with the latter lying consistently slightly 
lower than the corresponding POITSE values.
Within a rigid coupling approximation, the complete $^4$He density 
is assumed to adiabatically follow the rotational motion of OCS.  Therefore the 
systematically higher
values of the POITSE results is a direct manifestation of only 
{\em partial} 
adiabatic following of the solvating helium density with OCS rotation. 
The quasi-adiabatic rigid coupling approach does incorporate all zero point 
motions.
This results in the intersection at $N=4$ of the rigid coupling estimate of 
$B_{avg}$ with the 
experimental $B$ value in large droplets, in contrast to the crossing at 
$N$=6 
estimated in Ref.~\cite{grebenev00b} from a classical binding model that 
neglects zero 
point energy.  The actual crossing point is in remarkable agreement with the
integral of the molecule-induced local non-superfluid density 
calculated by path integral methods in Ref.~\cite{seekwon00}
($n_{ns} \sim 3.2$), and is thus consistent with the analysis made there of 
a reduced $B$ deriving from rigid coupling to the local non-superfluid density 
in the
first solvation shell. 
The quasi-adiabatic rigid coupling estimates
fail at larger sizes, where they necessarily continue to decrease monotonically, 
reaching
negligible values by $N \sim 50$.
Table~\ref{table_results} shows the quasi-adiabatic rigid coupling results for 
all three 
rotational 
constants $A, B$ and $C$, for selected cluster sizes. This indicates that the
asymmetric top spectrum ($A > B > C$) evolves into a prolate symmetric top 
spectrum 
($A > B=C$) by $N=10$, 
the same size at which the
POITSE $B_{avg}$
reaches the asymptotic droplet value of $B$ (Fig.~\ref{fig1}).
However, we expect that the transition between asymmetric and symmetric top may 
occur at 
a different cluster size for the exact calculations. 

One of the most interesting features of the POITSE results seen in 
Fig.~\ref{fig1}
is the turn-around of 
$B_{avg}$ at relatively small $N$, and subsequent rise to saturation at the 
experimental large droplet 
value by $N=10$.  
This behavior is related to the onset and nature of permutation exchanges 
between the 
$^4$He atoms. It can be quantitatively interpreted in terms of these 
exchanges~\cite{kwon02a}
and of the structures previously analyzed in terms of the axial solvation 
locations in Ref.~\cite{seepaesani01b}.
As shown there, while for $N\leq5$ the helium density is essentially
localized in the global minimum of the He-OCS potential energy surface, 
as $N$ increases beyond 5 the additional helium atoms solvate other regions 
along the molecule, covering the entire axial extent of 
the OCS molecule by $N=10$. 
Path integral calculations show no permutation exchanges
for $N \leq 5$,
but as $N$ increases above 5 and the helium density grows along the molecular
axis, exchange permutations are seen, with components both along and around
the molecular axis~\cite{kwon02a}. 
The lack of exchanges for $N \leq 5$ is attributed to the difficulty of
exchanging within a single tightly packed axial ring, and accounts for the near 
rigid behavior 
at these sizes.  The 
permutation exchange path components along the molecular axis seen for
$N \geq 6$ result in a lowering of the corresponding 
rigid body
response for rotation about axes perpendicular to the molecule ({\it i.e.},
to a non-zero perpendicular superfluid response), causing
in a rise in rotational constants $B$ and $C$~\cite{kwon02}. This 
effect continues 
as the solvation layer grows along the molecular axis and
the lateral permutation exchange contributions increase, thereby accounting 
for the 
observed increase in the POITSE values over the range $N=6 - 10$ in 
Fig.~\ref{fig1}.
Note that $N=10$ is the first size at which the $^4$He density extends along the 
entire molecular axis~\cite{seepaesani01b}, thereby allowing permutation 
exchanges of maximal extent within the first solvation shell, {\it i.e.}, from 
one end of the molecule to the other. 
We conclude that for OCS
the range $N=6-10$ with its increase of $B_{avg}$ to saturation 
from an overshoot at $N=6$ denotes a transition between a van der Waals 
(molecular) 
complex
and a true quantum solvated molecule.  The complex is 'near rigid', or 
'floppy',
in spectroscopic language, while the quantum solvated molecule 
is characterized by permutation exchanges between the helium atoms~\cite{seekwon00}. 
For large enough clusters, the latter leads to a microscopic local two-fluid 
description of co-existing superfluid and 
non-superfluid densities within the first solvation 
shell~\cite{kwon99,seekwon00}.   

In conclusion, we have provided calculation of rotational constants for
OCS-doped $^4$He clusters containing up to one complete solvation shell ($N = 1 
- 20$), 
by a direct method without imposing dynamical or nodal approximations.  We find 
accurate agreement with experimental
results at both extremes of small size $N=1, 2$, and large droplets.  
These POITSE calculations allow for the first time
the transition from a molecular complex to a quantum solvated molecule in 
superfluid $^4$He to be precisely identified, and interpreted in
terms of the quantum structure and permutation exchange propensity of the local 
helium environment.

{\em Acknowledgements:} 
This work was supported by the NSF (CHE-9616615, CHE-0107541) and by the Italian
Ministry for University and Research (MUIR). 
We thank NPACI and CASPUR for computation time at the San Diego Supercomputer 
Center and the University of Rome Computing Center, respectively.

\bibliographystyle{prsty}

\end{multicols}

\vspace{-0.5cm}
\begin{table}
\caption[]{Rotational constants (in cm$^{-1}$) calculated by the quasi-adiabatic rigid 
coupling method with two different He-OCS potential energy surfaces, shown for clusters 
having $N \leq 10$ $^4$He atoms. $A, B, C$: from HHDSD potential~\onlinecite{gianturco00b}.
$A', B', C'$: from MP4 potential~\cite{higgins99}.}
\label{table_results}
\begin{tabular}[t]{ccccccc}
$N$ & $A$ & $A'$ & $B$ & $B'$ & $C$ & $C'$ \\
\hline
     1  &  0.457(2) & 0.443(2) &  0.175(2) & 0.173(2) &  0.123(2) & 0.120(2) \\
     2  &  0.208(2) & 0.202(2) &  0.128(2) & 0.129(2) &  0.092(1) & 0.092(1) \\
     4  &  0.101(2) & 0.098(1) &  0.074(1) & 0.077(1) &  0.062(1) & 0.065(1) \\
     6  &  0.067(1) & 0.065(1) &  0.051(1) & 0.052(1) &  0.044(1) & 0.045(1) \\
     7  &  0.057(1) & 0.056(1) &  0.045(1) & 0.045(1) &  0.038(1) & 0.039(1) \\
    10  &  0.039(1) & 0.037(1) &  0.028(1) & 0.028(1) &  0.025(1) & 0.024(1) \\
\end{tabular}
\end{table}

\end{document}